\documentclass[cameraready]{Interspeech}

\usepackage{amsmath,amssymb,mathtools}
\usepackage{graphicx}
\usepackage[table]{xcolor}
\usepackage{booktabs}
\usepackage{adjustbox}
\usepackage{url}
\usepackage{float}
\usepackage{etoolbox}
\usepackage{algorithm2e}
\usepackage{comment}
\usepackage{subfig} 
\usepackage{hyperref}
\definecolor{rankpurple}{HTML}{a9def9}

\newcommand{\topyone}[1]{\cellcolor{rankpurple!180}\textbf{#1}}
\newcommand{\topytwo}[1]{\cellcolor{rankpurple!110}#1}
\newcommand{\topythree}[1]{\cellcolor{rankpurple!62}#1}

\title{From Signals to Patterns: Non-Invasive Tuberculosis Detection from Cough Audio using Bandit Weighted Hyperbolic Prototypes}

\author[affiliation={1}, orcid=0009-0000-1982-7110, equalcontribution]{Mohd Mujtaba}{Akhtar}
\author[affiliation={1}, orcid=0009-0004-2926-7777, equalcontribution]{Girish}{}
\author[affiliation={2}, orcid=0009-0003-3284-5008]{Sanjam}{Wadhwa}
\author[affiliation={1}, orcid=0009-0008-2638-7000, correspondingauthor]{Muskaan}{Singh}
\author[affiliation={3}, orcid=0000-0002-4112-3109]{Ning}{Ma}

\address{
$^1$ Ulster University, UK,
$^2$ Thapar Institute of Engineering and Technology, India,
$^3$ University of Sheffield, UK
}

\email{
girish.research.pr@gmail.com,
mmakhtar.research@gmail.com,
m.singh@ulster.ac.uk
}


\usepackage{comment}

\begin{document}

\maketitle

\begin{abstract}

\noindent In this study, we focus on cough-based tuberculosis screening (CBTS) and hypothesize that fusing speech/audio foundation representations with spectral descriptors will yield stronger screening performance. We expect this fusion to reveal complementary strengths: spectral features preserve fine-grained short-time acoustic detail in cough signals, while foundation embeddings capture higher-level temporal and event-level patterns learned from large-scale pretraining. To this end, we propose \textbf{\texttt{COBALT}}, a novel fusion framework based on codebook-aligned hyperbolic prototypes and bandit-style reliability weighting to integrate heterogeneous representations effectively. Using the CODA TB DREAM Challenge benchmark, \textbf{\texttt{COBALT}} consistently outperforms individual representations and a concatenation baseline, achieving the best overall performance when fusing MFCC with PaSST thereby establishing a new state-of-the-art on the benchmark.

\end{abstract}

\section{Introduction}
Tuberculosis (TB), a major infectious disease, often presents with persistent cough and other respiratory symptoms—signals that are not only clinically salient for screening but also reflect underlying pulmonary pathology and disease burden \cite{WHO_global_TB}. Accordingly, CBTS triage has attracted growing interest as a rapid, low-cost screening route; however, symptom-led pathways (often centered on cough) show highly variable diagnostic yield in practice and still require escalation to sputum-based confirmatory testing, which remains operationally demanding and difficult to scale uniformly across settings \cite{storla2008systematic,jiz243}. Recent advances in speech processing and self-supervised audio representation learning have shown promising capabilities in analyzing cough and other respiratory sounds for automated respiratory health inference \cite{niizumi25_interspeech}. 
However, translating these methods into practice is complicated by substantial cross-device and cross-environment variability, between-site population heterogeneity, and the need for careful pathway-level validation and integration within TB triage workflows \cite{jaganath2025accelerating,grant2022considerations}.

Early work primarily used handcrafted spectral descriptors with classical classifiers, establishing competitive baselines in clinical cohorts \cite{pahar2021automatic}. In this clinical screening task, attention-based sequence modeling has been applied to encode longer-range cough patterns and improve invariance \cite{frost22_interspeech}. As a step toward scalable, specimen-free triage, recent studies \cite{rajasekar2024detection} have investigated cough-audio modeling for CBTS—ranging from capsule-network classifiers trained on spectrogram-derived representations to speech foundation model–based systems studied in routine clinical workflows. In particular, leveraging pretrained models (PTMs) and attention-based sequence modeling of cough dynamics has enabled a more objective, data-driven characterization of cough patterns for TB-oriented screening \cite{AkhYas_NonInvasive_MICCAI2025}. As representation learning accelerates progress in CBTS, \textit{the central challenge is to infer tuberculosis from cough acoustics under substantial real-world variability, and to determine whether higher-level representations—typically learned from large-scale speech signals—remain informative when the input contains little or no linguistic content.} Rigorous validation of CBTS has become vital for public-health case finding, yet understanding what cues models truly rely on—versus device or environment artifacts—remains an open challenge. While CBTS from audio has seen notable progress \cite{sharma2024tbscreen}, the equally critical challenge of understanding what cues models truly rely on—rather than device or environment artifacts—remains largely under-explored. However, the community has largely focused on improving single-stream backbones, while systematic treatment of cross-representation complementarity—and how to exploit it through principled integration remains sparse. In this work, for the first time and to the best of our knowledge, we systematically study the fusion of heterogeneous CBTS representations—foundation-model embeddings and classical spectral descriptors. \textit{We hypothesize that fusing these complementary views will yield stronger tuberculosis screening performance, since spectral descriptors preserve fine-grained acoustic detail (e.g., short-time spectral structure), while foundation embeddings capture higher-level temporal and event-level patterns in cough audio.} To validate our hypothesis and support effective fusion, we propose a novel framework, \textbf{\texttt{COBALT}} — \underline{CO}debook-Aligned \underline{BA}ndit-weighted hyperbo\underline{L}ic pro\underline{T}otypE fusion, which unifies heterogeneous pretrained streams by mapping their adapted tokens into a hyperbolic space and aligning them through a shared prototype codebook via hyperbolic vector quantization. As far as we know, this is the first study to couple codebook-aligned hyperbolic prototype learning with a bandit-style reliability mechanism for fusion in CBTS. By quantizing both streams into the same hyperbolic prototype vocabulary and learning reliability weights over prototypes, \textbf{\texttt{COBALT}} preserves complementary cues across time–frequency scales while down-weighting unstable, artifact-prone evidence, yielding a more coherent and informative fused representation.

\begin{figure*}[!hbt]
    \centering
    \includegraphics[width=0.8\linewidth]{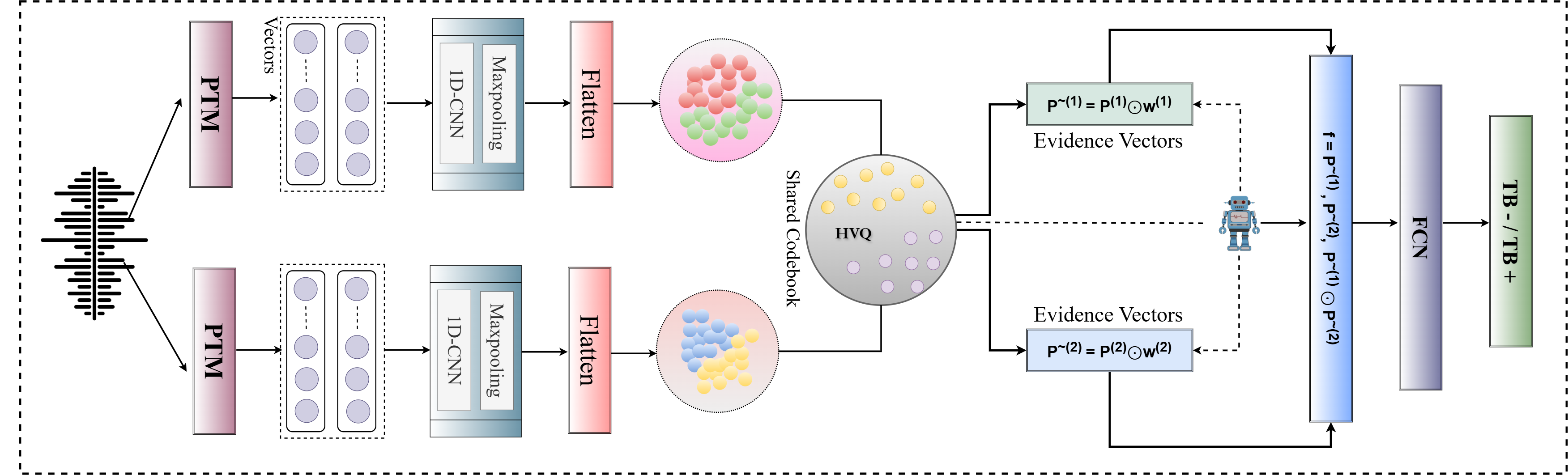}
    \caption{Proposed \textbf{\texttt{COBALT}} framework for cough-based tuberculosis screening.}
    \label{fig:proposed}
\end{figure*}

\noindent \textbf{The key contributions of this work are:}
\begin{itemize}
\item We present a comprehensive benchmark for cough-based tuberculosis screening (CBTS), evaluating diverse pretrained foundation representations alongside classical spectral descriptors under a unified protocol.
\item We propose \textbf{\texttt{COBALT}} (Figure~\ref{fig:proposed}), a novel fusion framework that aligns heterogeneous pretrained streams via a shared hyperbolic prototype codebook and learns bandit-based reliability weights to perform effective prototypE fusion.
\item Using \textbf{\texttt{COBALT}}, we achieve superior performance compared to individual representations and baseline fusion methods, demonstrating the benefit of principled multi-representation fusion for CBTS.
\end{itemize}
\footnote{Project resources are publicly available at: \url{https://github.com/Helixometry/COBALT.git}.}

\section{Representations}
In this section, we outline the pretrained representation models and handcrafted spectral features used in our experiments. \newline
\noindent \textbf{Pretrained representation models:}
PaSST\footnote{\url{https://github.com/kkoutini/PaSST}} \cite{koutini22_interspeech} is a spectrogram Transformer that adapts ViT-style patch tokenization to audio and employs patch masking for regularization; we use the PaSST-S variant (87M). Whisper\footnote{\url{https://huggingface.co/openai/whisper-base}} \cite{radford2023robust} is a multilingual ASR model trained at scale on weakly supervised audio--text data; we adopt the Base checkpoint (74M) and extract encoder embeddings as a general-purpose speech representation. x-vector\footnote{\url{https://huggingface.co/speechbrain/spkrec-xvect-voxceleb}} \cite{8461375} is a compact TDNN-based speaker embedding model (4.2M) trained for speaker recognition, which we include as a lightweight pretrained baseline. WavLM\footnote{\url{https://huggingface.co/microsoft/wavlm-base}} \cite{chen2022wavlm} is a self-supervised speech representation model that extends masked prediction with mixture-based objectives; we use the Base variant (94M) pretrained on 94k hours of audio. \newline
\noindent For all pretrained models, we keep the encoders frozen and extract representations from the final hidden layer using average pooling to obtain a fixed-dimensional utterance embedding. The resulting embedding dimensions are 768 for PaSST, WavLM, and Whisper (Base), and 512 for x-vector.

\noindent \textbf{Handcrafted spectral features:}
In addition to pretrained embeddings, we extract MFCC and LFCC as standard handcrafted spectral descriptors. Both are computed on short-time frames using a Mel- or linear-frequency filterbank, followed by a DCT to obtain cepstral coefficients. For MFCCs, we set the cepstral dimensionality to 40. We then apply statistics pooling (mean and standard deviation) over time to produce a fixed-length representation for classification. MFCCs are extracted using \texttt{librosa}\footnote{\url{https://librosa.org/doc/main/generated/librosa.feature.mfcc.html}}, while LFCCs are computed using \texttt{spafe}\footnote{\url{https://spafe.readthedocs.io/en/latest/features/lfcc.html}}.

\section{Modeling}

\subsection{Proposed framework : \textbf{\texttt{COBALT}}}

We propose \textbf{\texttt{COBALT}} (\textit{Codebook-Aligned Bandit-weighted prototypE fusion Learning}) for fusing dual pretrained-model (PTM) representations for cough-based tuberculosis detection. The overall architecture of \textbf{\texttt{COBALT}} is shown in Figure~\ref{fig:proposed}.

\noindent \textbf{Dual-stream representation encoding:} Let $x$ denote an input cough sample and $B$ the batch size. We extract two sequence representations using two PTMs:
\begin{equation*}
\begin{adjustbox}{max width=\linewidth}
$\aligned
X^{(1)}=\mathrm{PTM}_1(x)\in\mathbb{R}^{B\times T_1\times d_1},\qquad
X^{(2)}=\mathrm{PTM}_2(x)\in\mathbb{R}^{B\times T_2\times d_2}
\endaligned$
\end{adjustbox}
\end{equation*}
\textbf{\texttt{COBALT}} compresses each stream into $K$ tokens ($K\ll T_1,T_2$) and learns a shared codebook of $M$ prototypes in a $d_h$-dimensional hyperbolic space. We use the Poincar\'e ball model $\mathbb{B}^{d_h}_c$ with curvature $c>0$.

\noindent \textbf{Stream adaptation and tokenization:}
We first adapt each PTM stream to the target task and align their temporal resolution and dimensionality using lightweight 1D CNN adapters $g_m(\cdot)$ (projection with pooling/striding):
\begin{equation*}
H^{(m)}=g_m(X^{(m)})\in\mathbb{R}^{B\times T^{\prime}\times d},\quad m\in\{1,2\}
\end{equation*}
We then compress each adapted sequence into $K$ tokens via a tokenization operator $\mathrm{Tokenize}(\cdot)$ (e.g., adaptive pooling or learnable attention pooling):
\begin{equation*}
Z^{(m)}=\mathrm{Tokenize}(H^{(m)})\in\mathbb{R}^{B\times K\times d},\quad m\in\{1,2\}
\end{equation*}

\noindent \textbf{Hyperbolic mapping and prototype alignment:}
Next, we project the $K$ tokens from each stream into a $d_h$-dimensional latent space and map them to the Poincar\'e ball $\mathbb{B}^{d_h}_c$ using the exponential map at the origin:
\begin{equation*}
Y^{(m)}=\exp_{0}^{c}(W_m Z^{(m)})\in\mathbb{B}^{d_h}_c,\quad m\in\{1,2\}
\end{equation*}
We maintain a \textbf{shared} hyperbolic codebook $\mathcal{C}=\{c_1,\dots,c_M\}$, $c_j\in\mathbb{B}^{d_h}_c$, and softly assign each token to prototypes based on hyperbolic distance:
\begin{equation*}
A_j(y)=
\frac{\exp(-d_{\mathbb{B}}(y,c_j)/\tau_q)}
{\sum_{\ell=1}^{M}\exp(-d_{\mathbb{B}}(y,c_{\ell})/\tau_q)}
\end{equation*}
This shared codebook aligns both streams into a common prototype vocabulary for subsequent fusion.

\noindent \textbf{Prototype evidence and reliability weighting:}
We summarize each stream by aggregating token-to-prototype assignments into a prototype evidence vector $p^{(m)}\in\mathbb{R}^{B\times M}$:
\begin{equation*}
p^{(m)}_{bj}=\frac{1}{K}\sum_{k=1}^{K}A^{(m)}_{bkj},\quad m\in\{1,2\}
\end{equation*}
We then learn global prototype reliabilities using a multi-armed bandit view, where each prototype $j$ maintains a score $Q_j$ and induces a weight vector
\begin{equation*}
w=\mathrm{softmax}(Q/\tau_w).
\end{equation*}
Using these weights, we reweight the per-stream evidence as $\tilde{p}^{(m)}=p^{(m)}\odot w$, emphasizing informative prototypes and suppressing unreliable evidence during fusion.

\noindent \textbf{Bandit reward and update:}
Next, we compute a reward that reflects the benefit of reliability weighting relative to a baseline. Let $L_{\mathrm{base}}$ denote a baseline loss (e.g., uniform weights or a moving-average baseline) and $L_{\mathrm{COBALT}}$ the loss obtained using $\tilde{p}^{(m)}$; we define
\begin{equation*}
r=\alpha\left(L_{\mathrm{base}}-L_{\mathrm{COBALT}}\right)+\beta\left(M_{\mathrm{COBALT}}-M_{\mathrm{base}}\right),
\end{equation*}
where $M$ is a confidence margin and $\alpha,\beta\ge0$. We then update each prototype score using a usage-weighted exponential moving rule,
\begin{equation*}
Q_j \leftarrow (1-\eta u_j)Q_j + (\eta u_j) r,
\end{equation*}
with step size $\eta>0$ and batch usage $u_j$ computed from the current prototype assignments.

\noindent \textbf{PrototypE fusion and prediction:}
Finally, we form the fused representation by concatenating the two reweighted evidence vectors and their agreement:
\begin{equation*}
f=\Big[\tilde{p}^{(1)},\ \tilde{p}^{(2)},\ \tilde{p}^{(1)}\odot \tilde{p}^{(2)}\Big],
\end{equation*}
and feed $f$ to a lightweight MLP classifier to obtain the final prediction $\hat{y}=\mathrm{MLP}(f)$

\noindent \textbf{Training objective:}
We train COBALT end-to-end using a task loss (cross-entropy) together with HVQ regularization:
\begin{equation*}
\mathcal{L}
=
\mathcal{L}_{\mathrm{task}}(\hat{y},y)
+
\beta_{\mathrm{vq}}
(\mathcal{L}^{(1)}_{\mathrm{vq}}+\mathcal{L}^{(2)}_{\mathrm{vq}})
+
\lambda\,H(w)
\end{equation*}

\noindent where $\mathcal{L}^{(m)}_{\mathrm{vq}}$ denotes the stream-wise vector-quantization loss terms and $H(w)=-\sum_{j=1}^{M}w_j\log w_j$ encourages selective prototype usage. We ensure that the mapped hyperbolic tokens remain inside the Poincar\'e ball (i.e., $\lVert y\rVert < 1$) to maintain numerical stability. We attach a lightweight MLP/FCN head on top of the fused representation $f$, consisting of a dense layer (e.g., 128 units) followed by a softmax output layer that produces class probabilities. The number of trainable parameters depends on the selected PTM pair and adapter dimensions, ranging from \textit{3}M to \textit{6}M in our implementations.

\section{Experiment}

\subsection{Benchmark Dataset}
\label{corpus}
Our study is grounded in experiments on the CODA TB\cite{huddart2024dataset}\footnote{\url{https://www.synapse.org/Synapse:syn31472953/wiki/619711}} benchmark dataset released through the CODA TB DREAM Challenge. It consists of solicited cough audio from adult participants ($\geq$18 years) presenting with respiratory symptoms, collected at outpatient health centers spanning seven countries (India, Madagascar, the Philippines, South Africa, Tanzania, Uganda, and Vietnam). We follow the official subject-disjoint splits provided with the benchmark to ensure no participant overlap between training and evaluation partitions. \newline
\noindent\textbf{Training and Hyperparameter Details:}
We train all models for 50 epochs with a batch size of 32, optimizing cross-entropy using Adam. Performance is reported under five-fold cross-validation, using four folds for training and one fold for evaluation per split.

\begin{table}[!hbt]
\centering
\begin{adjustbox}{width=\linewidth,center}
\begin{tabular}{l|ccc|ccc}
\toprule
& \multicolumn{3}{c|}{\textbf{FCN}} & \multicolumn{3}{c}{\textbf{CNN}} \\
\cmidrule(lr){2-4}\cmidrule(lr){5-7}
\textbf{PTM's} & \textbf{ACC \(\uparrow\)} & \textbf{F1 \(\uparrow\)} & \textbf{AUC \(\uparrow\)} & \textbf{ACC \(\uparrow\)} & \textbf{F1 \(\uparrow\)} & \textbf{AUC \(\uparrow\)} \\
\midrule
WHS  & 73.24 & 71.82 & 68.46 & 76.56 & 75.12 & 70.23 \\
WAL  & 69.87 & 68.15 & 58.62 & 72.09 & 70.46 & 64.57 \\
XCR  & \topythree{74.01} & \topythree{73.67} & \topythree{69.51} & \topythree{77.81} & \topythree{76.03} & \topythree{71.82} \\
MF   & \topytwo{75.62} & \topytwo{74.29} & \topytwo{70.18} & \topytwo{77.32} & \topytwo{75.69} & \topytwo{70.36} \\
LF   & 73.43 & 71.96 & 65.93 & 75.93 & 74.28 & 69.14 \\
PST  & \topyone{78.92} & \topyone{77.61} & \topyone{72.20} & \topyone{79.29} & \topyone{77.57} & \topyone{72.68} \\
\bottomrule
\end{tabular}
\end{adjustbox}
\caption{Performance scores(\%) of individual representations with FCN and CNN downstream models; Abbreviations used: WHS (Whisper), WAL (WavLM), XCR (x-vector), PST (PaSST), MF (MFCC), and LF (LFCC). Scores are averaged over five folds; the same abbreviations are used in Table~\ref{tab:fusion}.}
\label{tab:baseline}
\vspace{-25pt}
\end{table}

\subsection{Experimental Results}

We report the results of our experiments with individual pre-trained representations (PTMs) and spectral features (SFs) in Table~\ref{tab:baseline}. Performance is evaluated using Accuracy (ACC), F1-score (F1), and Area Under the ROC Curve (AUC). Table~\ref{tab:baseline} shows that PaSST (PST) delivers the strongest standalone performance with both backbones, and that the CNN backend generally provides a consistent improvement over FCN across representation types. Among PTMs, PST consistently outperforms Whisper (WHS), WavLM (WAL), and x-vector (XCR) under both FCN and CNN, which we attribute to its spectrogram-based transformer pretraining that better preserves cough-relevant time–frequency structure. Among handcrafted SFs, MFCC emerges as the strongest classical baseline (followed by LFCC), remaining competitive with the mid-tier PTMs across metrics and benefiting notably from the CNN backend. We also observe mild shifts in the relative ranking of representations between FCN and CNN, suggesting that downstream inductive bias can influence how effectively each feature family is leveraged. Finally, these trends align with recent findings from high-burden settings showing that cough analysis with SFMs can yield strong screening performance, further supporting the utility of pretrained representations for CBTS \cite{ma2025deeplearni}.

\begin{table}[!hbt]
\centering
\begin{adjustbox}{width=\linewidth,center}
\begin{tabular}{l|ccc|ccc}
\toprule
& \multicolumn{3}{c|}{\textbf{Concate}} & \multicolumn{3}{c}{\textbf{COBALT-E}} \\
\cmidrule(lr){2-4}\cmidrule(lr){5-7}
\textbf{Pairs} & \textbf{ACC \(\uparrow\)} & \textbf{F1 \(\uparrow\)} & \textbf{AUC \(\uparrow\)} & \textbf{ACC \(\uparrow\)} & \textbf{F1 \(\uparrow\)} & \textbf{AUC \(\uparrow\)} \\
\midrule
WHS + WAL  & 78.91 & 77.23 & 74.58 & 82.37 & 80.72 & 78.43 \\
WHS + XCR  & 80.04 & 78.69 & 76.13 & 83.29 & 82.01 & 81.69 \\
WHS + LF   & 78.62 & 76.91 & 73.84 & 81.92 & 80.39 & 78.04 \\
WHS + MF   & 79.58 & 78.02 & 76.29 & 82.74 & 81.27 & 80.56 \\
WHS + PST  & \topythree{81.47} & 79.54 & \topytwo{80.61} & \topythree{84.65} & \topythree{82.94} & 81.37 \\
WAL + XCR  & 80.19 & 78.36 & 78.92 & 83.56 & 81.63 & \topythree{82.96} \\
WAL + LF   & 78.31 & 77.20 & 75.06 & 81.94 & 80.69 & 78.13 \\
WAL + MF   & 79.46 & 78.15 & 74.39 & 82.78 & 81.45 & 79.48 \\
WAL + PST  & 80.94 & 79.68 & 79.52 & 84.13 & 82.70 & 81.29 \\
XCR + LF   & 79.11 & 77.82 & 76.23 & 82.47 & 81.36 & 80.76 \\
XCR + MF   & 80.23 & 79.06 & 78.94 & 83.61 & 82.05 & \topytwo{83.14} \\
XCR + PST  & \topyone{81.74} & \topytwo{80.27} & \topythree{79.87} & \topytwo{84.19} & \topytwo{83.24} & 80.57 \\
LF + MF    & 79.58 & 77.96 & 75.02 & 82.74 & 81.47 & 80.91 \\
LF + PST   & 81.42 & \topyone{80.39} & 79.44 & 84.58 & 82.61 & 82.34 \\
MF + PST   & \topytwo{81.67} & \topythree{79.88} & \topyone{81.27} & \topyone{85.97} & \topyone{83.54} & \topyone{85.06} \\
\bottomrule
\end{tabular}
\end{adjustbox}

\caption{Performance scores (\%) comparing simple concatenation baseline and the Euclidean variant \textbf{\texttt{COBALT-E}}.}
\label{tab:2}
\vspace{-25pt}
\end{table}

\noindent Table~\ref{tab:2} compares a naïve fusion baseline based on feature concatenation (Concate) with a Euclidean instantiation of our fusion pipeline, \textbf{\texttt{COBALT-E}}. This experiment is intended as a geometry-control: it keeps the fusion principle intact while removing hyperbolic operations, allowing us to assess how much structure can be recovered even in a purely Euclidean setting. Across all representation pairs, \textbf{\texttt{COBALT-E}} yields consistent improvements over concatenation in ACC, F1, and AUC, indicating that simply stacking embeddings is suboptimal and that learning a structured fusion function is beneficial. The strongest concatenation results are typically obtained when PST participates in the pair (e.g., WHS+PST and MF+PST), suggesting that PST provides a highly informative anchor representation. Under \textbf{\texttt{COBALT-E}}, MF+PST emerges as the strongest pairing, achieving 85.97 ACC, 83.54 F1, and 85.06 AUC. Notably, improvements are also evident for fusion (e.g., WAL+XCR) and SF–SF (LF+MF) combinations, highlighting that the benefit is not restricted to any single feature family but reflects a general advantage of guided fusion over naïve concatenation.

\begin{figure}[!hbt]
    \centering
    \subfloat[]{%
        \includegraphics[width=0.2\textwidth]{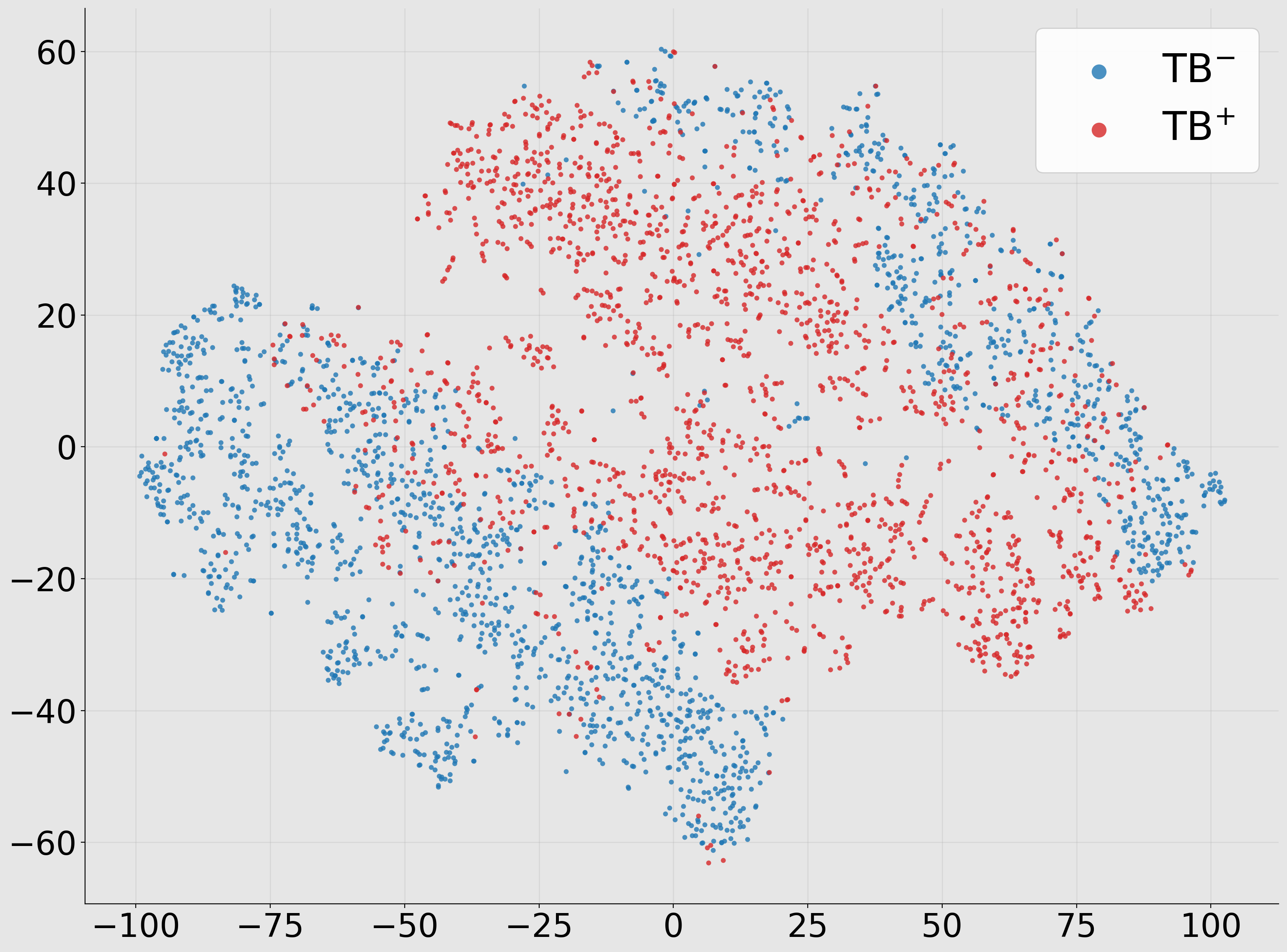}
    }
    \hspace{0.3mm}
    \subfloat[]{%
        \includegraphics[width=0.2\textwidth]{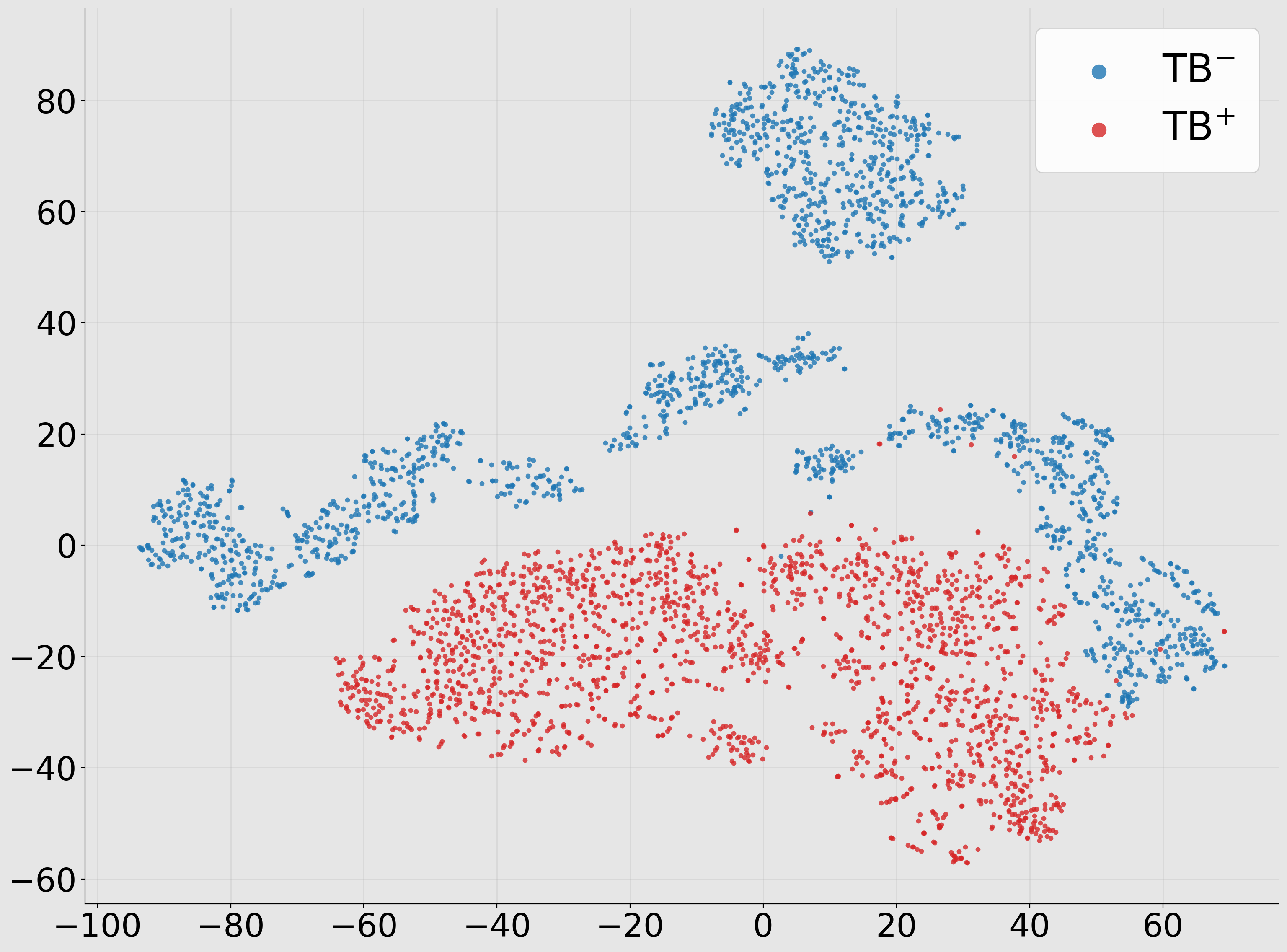}
    }

    \caption{t-SNE Plots for - (a) X-vector (b) MFCC+PaSST (\textbf{\texttt{COBALT}}) }
    \label{fig:tsne}
    \vspace{-20pt}
\end{figure}

\begin{figure}[!hbt]
    \centering
    \subfloat[]{%
        \includegraphics[width=0.23\textwidth]{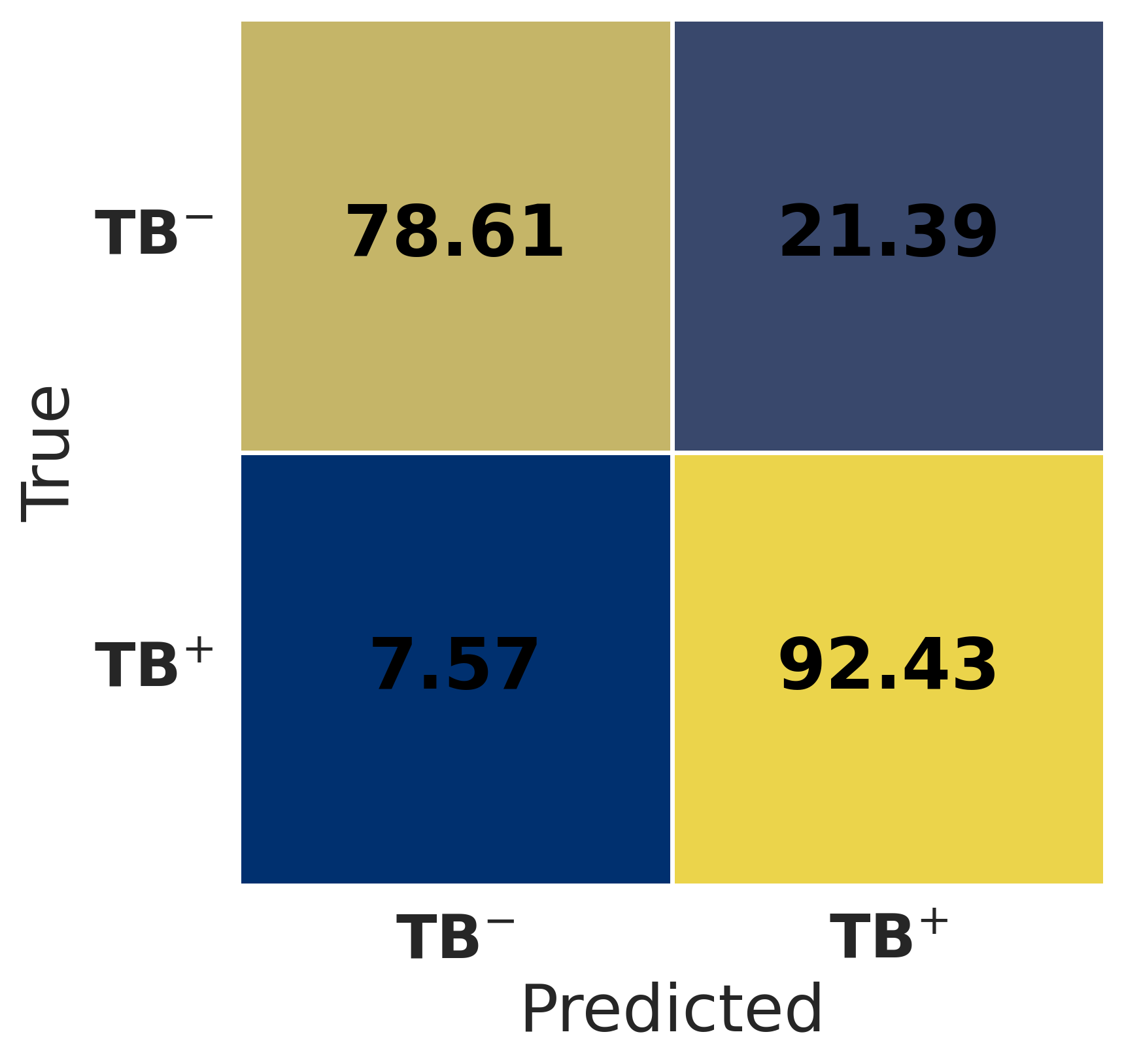}
    }
    \hfill
    \subfloat[]{%
        \includegraphics[width=0.23\textwidth]{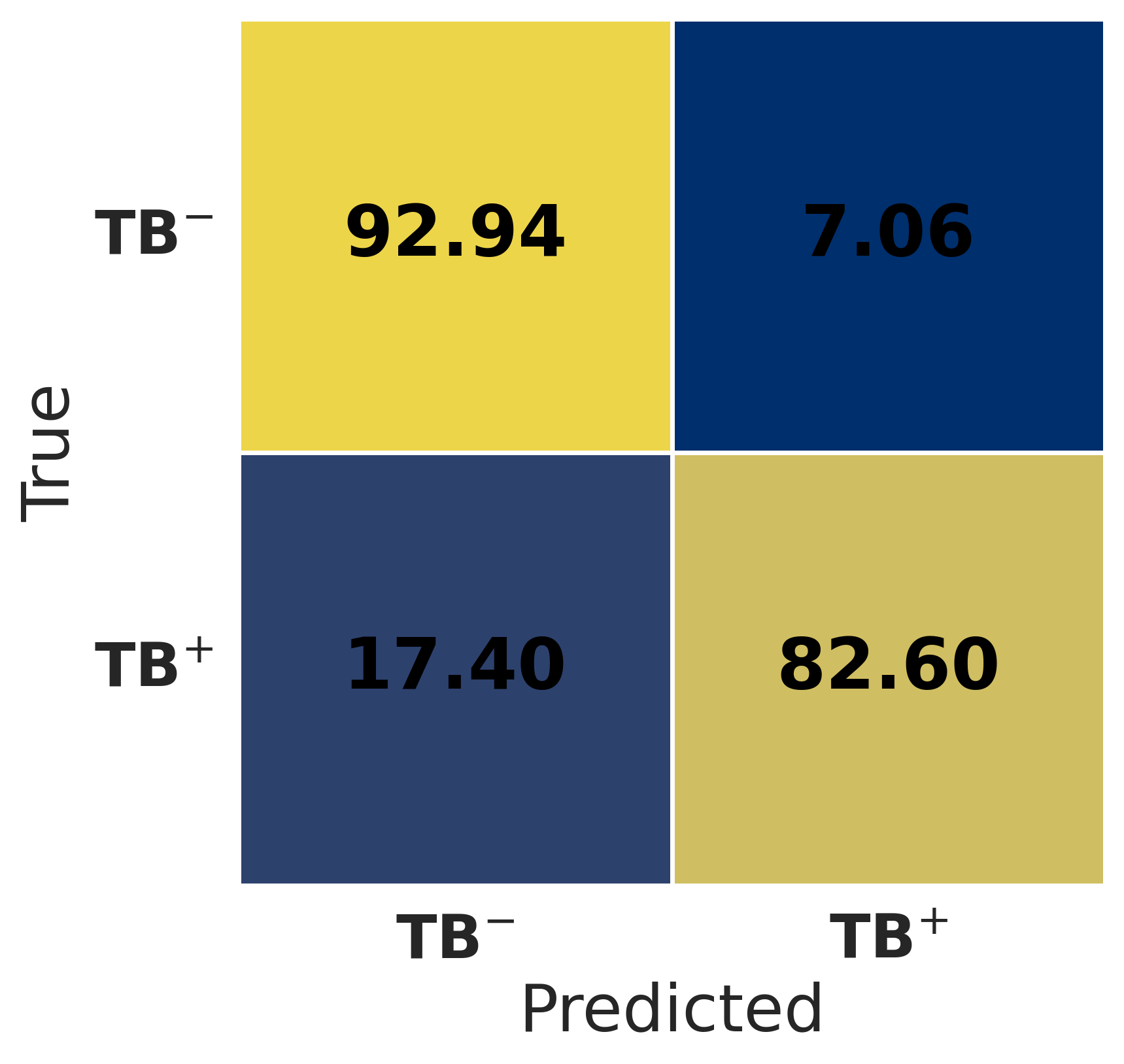}
    }

    \caption{Confusion Matrices - (a) Whisper+WavLM (Concat), (b) MFCC+PaSST (\textbf{\texttt{COBALT}}) }
    \label{fig:cm}
    \vspace{-15pt}
\end{figure}

\begin{table}[!hbt]
\centering
\begin{adjustbox}{width=\linewidth,center}
\begin{tabular}{l|ccc|ccc}
\toprule
& \multicolumn{3}{c|}{\textbf{Möbius addition}} & \multicolumn{3}{c}{\textbf{COBALT}} \\
\cmidrule(lr){2-4}\cmidrule(lr){5-7}
\textbf{Pairs} & \textbf{ACC \(\uparrow\)} & \textbf{F1 \(\uparrow\)} & \textbf{AUC \(\uparrow\)} & \textbf{ACC \(\uparrow\)} & \textbf{F1 \(\uparrow\)} & \textbf{AUC \(\uparrow\)} \\
\midrule
WHS + WAL  & 84.19 & 82.53 & 81.04 & 86.57 & 84.38 & 83.26 \\
WHS + XCR  & 82.34 & 81.68 & 79.52 & 85.14 & 83.62 & 80.69 \\
WHS + LF   & 79.98 & 78.23 & 76.81 & 82.92 & 80.39 & 78.04 \\
WHS + MF   & 83.52 & 82.17 & 80.96 & 86.47 & 85.24 & 83.79 \\
WHS + PST  & 85.79 & 84.32 & \topythree{83.68} & 87.65 & 85.94 & 83.37 \\
WAL + XCR  & 82.26 & 80.47 & 81.93 & 85.09 & 83.78 & 82.16 \\
WAL + LF   & 83.54 & 82.29 & 80.71 & 85.97 & 84.35 & 83.02 \\
WAL + MF   & 81.67 & 80.15 & 78.38 & 84.38 & 82.64 & 80.78 \\
WAL + PST  & \topyone{86.92} & \topytwo{85.49} & \topytwo{84.27} & \topytwo{88.26} & \topyone{87.52} & \topytwo{86.11} \\
XCR + LF   & 83.51 & 82.13 & 80.92 & 85.67 & 84.39 & 82.94 \\
XCR + MF   & 82.43 & 81.29 & 82.05 & 84.78 & 82.46 & 81.16 \\
XCR + PST  & \topytwo{86.29} & \topythree{84.61} & 81.97 & \topythree{88.02} & \topytwo{87.39} & \topythree{85.63} \\
LF + MF    & 81.76 & 80.13 & 78.65 & 84.19 & 82.98 & 81.45 \\
LF + PST   & 85.93 & 83.47 & 82.97 & 87.56 & 86.47 & 85.34 \\
MF + PST   & \topythree{86.05} & \topyone{85.69} & \topyone{86.78} & \topyone{88.93} & \topythree{87.26} & \topyone{89.07} \\
\bottomrule
\end{tabular}
\end{adjustbox}
\caption{Performance scores (\%) comparing Möbius-addition fusion and the proposed \textbf{\texttt{COBALT}} framework.}
\label{tab:fusion}
\vspace{-20pt}
\end{table}
\noindent Table~\ref{tab:fusion} summarizes a hyperbolic fusion ablation in which paired representations are composed using Möbius addition. By disabling the prototype codebook and the bandit-based reliability weighting, this configuration isolates the contribution of hyperbolic composition, serving as a controlled comparison to the full \textbf{\texttt{COBALT}} framework. For a fair comparison, we keep the downstream modeling and training protocol identical to the full \textbf{\texttt{COBALT}} setting, and modify only the fusion mechanism. The results show that Möbius-addition fusion consistently improves upon naïve concatenation across ACC, F1, and AUC, indicating that geometry-aware composition provides a stronger inductive bias for integrating heterogeneous cough representations than direct feature stacking. At the same time, the remaining gap to the complete \textbf{\texttt{COBALT}} model highlights the contribution of the two removed modules: the prototype codebook promotes a shared alignment between mismatched embedding spaces, while the bandit-based weighting adjusts each stream’s contribution in a sample-dependent manner, limiting the effect of less informative features. Finally, performance trends indicate that heterogeneous fusion is generally the most effective, with pairs involving strong single-stream features such as PST and MF typically ranking among the top configurations, reinforcing their complementary strengths for cough-based TB screening. These results support \textit{our hypothesis that combining complementary views leads to stronger tuberculosis screening performance: spectral descriptors retain fine-grained acoustic detail (e.g., short-time time--frequency structure), while foundation embeddings encode broader temporal context and event-level patterns in cough audio.} Moreover, \textbf{\texttt{COBALT}} enables effective integration of foundation and spectral representations, surpassing the strongest individual encoders in CBTS and providing consistent improvements over both homogeneous PTM pairings and the concatenation baseline. To further illustrate the improved class separability, we visualize the fused embeddings using t-SNE in Figure~\ref{fig:tsne} and report the corresponding confusion matrices in Figure~\ref{fig:cm}.

\section{Conclusion}

In this study, we present a systematic benchmark for cough-based tuberculosis screening (CBTS) and show that large pretrained acoustic encoders are strong representation learners for this task, with PaSST emerging as the most effective single encoder in our setting. At the same time, spectral features remain competitive and provide complementary information, motivating heterogeneous fusion. Building on these insights, we propose \texttt{\textbf{COBALT}}, a novel fusion framework that aligns heterogeneous representations via a shared hyperbolic prototype codebook and integrates them using bandit-style reliability weighting. \texttt{\textbf{COBALT}} consistently outperforms individual representations and a concatenation baseline, with the best results obtained by fusing spectral descriptors with foundation embeddings. Our work also calls upon researchers to extend our benchmark setting and continue improving CBTS performance.

\section{Acknowledgements}

The authors gratefully acknowledge the support of the United States--Ireland--Northern Ireland R\&D Partnership Programme (USI-207), and access to the Tier 2 High-Performance Computing resources from the Northern Ireland High Performance Computing (NI-HPC) service funded by EPSRC (EP/T022175).


\section{Generative AI Use Disclosure}
AI assistants were used only to refine grammar, enhance clarity, and improve the manuscript’s overall readability and presentation. They did not contribute to the development of scientific ideas, the design or execution of analyses, the production of results, or the interpretation of findings. The authors assume full responsibility for the accuracy and integrity of the work.

\bibliographystyle{IEEEtran}
\bibliography{mybib}

\end{document}